\documentclass[twocolumn,aps,pra,superscriptaddress]{revtex4-2}
\usepackage{amsmath}
\usepackage{enumitem}
\usepackage{amssymb}
\usepackage{bm}
\usepackage[utf8]{inputenc}
\usepackage{graphicx} 
\usepackage{dcolumn} 
\usepackage{wrapfig}
\usepackage[table, dvipsnames]{xcolor}
\usepackage[english]{babel}
\usepackage[unicode=true, colorlinks=true, citecolor={blue!80!black}, urlcolor={blue!50!black}, linkcolor = {blue!80!black}]{hyperref}
\usepackage{siunitx}
\usepackage{easyReview}
\usepackage{adjustbox}
\usepackage[T1]{fontenc}
\usepackage{blindtext}
\usepackage{braket}

\begin{document}

\title{Automated, physics-guided, multi-parameter design optimization for superconducting quantum devices}

\author{Axel M. Eriksson}
\email[Correspondence to: ]{axel.eriksson@chalmers.se}
\affiliation{Department of Microtechnology and Nanoscience, Chalmers University of Technology, 412 96 Gothenburg, Sweden}
\author{Lukas J. Splitthoff}
\affiliation{Department of Microtechnology and Nanoscience, Chalmers University of Technology, 412 96 Gothenburg, Sweden}
\author{Harsh Vardhan Upadhyay}
\affiliation{Department of Microtechnology and Nanoscience, Chalmers University of Technology, 412 96 Gothenburg, Sweden}
\author{Pietro Campana}
\affiliation{Department of Microtechnology and Nanoscience, Chalmers University of Technology, 412 96 Gothenburg, Sweden}
\affiliation{Department of Physics, University of Milano-Bicocca, 20126 Milan, Italy}
\author{Niranjan Pittan Narendiran}
\affiliation{Department of Microtechnology and Nanoscience, Chalmers University of Technology, 412 96 Gothenburg, Sweden}
\author{Kunal Helambe}
\affiliation{Department of Microtechnology and Nanoscience, Chalmers University of Technology, 412 96 Gothenburg, Sweden}
\author{Linus Andersson}
\affiliation{Department of Microtechnology and Nanoscience, Chalmers University of Technology, 412 96 Gothenburg, Sweden}
\author{Simone Gasparinetti}
\email[Correspondence to: ]{simone.gasparinetti@chalmers.se}
\affiliation{Department of Microtechnology and Nanoscience, Chalmers University of Technology, 412 96 Gothenburg, Sweden}

\begin{abstract}
The design of nonlinear superconducting quantum circuits often relies on time-consuming iterative electromagnetic simulations requiring manual intervention. These interventions entail, for example, adjusting design variables such as resonator lengths or Josephson junction energies to meet target parameters such as mode frequencies, decay rates, and coupling strengths. Here, we present a method to efficiently automate the optimization of superconducting circuits, which significantly reduces the need for manual intervention. The method's efficiency arises from user-defined, physics-informed, nonlinear models that guide parameter updates toward the desired targets. Additionally, we provide a full implementation of our optimization method as an open-source Python package, QDesignOptimizer. The package automates the design workflow by combining high-accuracy electromagnetic simulations in Ansys HFSS and Energy Participation Ratio (pyEPR) analysis integrated with the design tool Qiskit-Metal. Our implementation supports modular and flexible subsystem-level analysis and is easily extensible to optimize for additional parameters. The method is not specific to superconducting circuits; as such, it can be applied to a range of nonlinear optimization problems across science and technology. 
\end{abstract}
\maketitle

\section{Introduction}
Nonlinear superconducting circuits, built around Josephson junctions, are a central building block for superconducting quantum processors \cite{Blais_2004, Krantz_2019, Bravyi_2024, Acharya2025} and parametric amplifiers \cite{Castellanos_2007}.
The process of translating an initial concept of a superconducting circuit (\textit{circuit-level design}) into a finalized physical implementation (which requires a \textit{layout-level design}) is inherently complex, iterative, and time-consuming \cite{Krantz_2019, Gao_2021, Levenson_2024}. 

The circuit design process usually begin with a user-defined circuit Hamiltonian, which determines all target physical quantities, such as mode frequencies, nonlinear couplings and decay rates, which lead to the intended functionality. The target circuit is then mapped onto a physical layout composed of lumped and distributed circuit elements. Common design platforms for this task include IBM’s Qiskit-Metal \cite{Minev_2021}, IQM’s KQCircuits \cite{kqcircuits}, and gdsfactory \cite{gdsfactory_git}.

For initial approximate circuit designs, lumped-element circuits are efficiently analyzed using simplified circuit models and fast component-wise simulations \cite{Gely_2020, Groszkowski_2021, Tanamoto_2023, Kunasaikaran_2024}. However, accounting for the distributed nature of components requires more sophisticated eigenmode, capacitance, or scattering simulations, for example using open-source\cite{CSC_2023} or commercial simulation software \cite{ansys_2021, Sonnet_2023}. These high-accuracy simulations are time-consuming and computationally intensive but provide high predictive power, especially needed for distributed circuits.

To converge on a final design, some update strategy must be applied which iteratively brings the design closer to the target. The update strategy is often manual, which slows down the design process. 

To accelerate the design process, many research groups working on superconducting circuits have established their internal workflows for layouts and simulations, often tailored to their device scale and complexity. One notable effort towards open-source optimization of physical layouts has led to the automation of superconducting coupler architecture designs \cite{Li_2023}. Another, open-source contribution is the {\it SQuADDS} database of superconducting quantum devices serving as a starting point for customized devices \cite{Shanto_2024}. However, these contributions are limited to specific circuit geometries, which have been studied or pre-simulated making them difficult to extend to novel designs.

Here, we present a method which automates the manual update step through an efficient, physics-guided, multi-parameter optimization framework. The user supplies their physical understanding of the circuit as an approximate custom nonlinear model to guide design variables updates. The framework automates the optimization for arbitrary device geometries, naturally incorporating parameter inter-dependencies and supporting easy extension to new models and targets. Our method integrates with Qiskit-Metal-based designs \cite{Minev_2021} and high-accuracy electromagnetic simulations in Ansys HFSS \cite{ansys_2021}, which are further analyzed with energy-participation ratio studies in pyEPR \cite{Minev_2021_epr}. The underlying method of solving an approximate nonlinear model for fast convergence can be applied to any computationally heavy, multidimensional optimization task, making it relevant for a vast range of engineering problems. The optimization is feasible under two main conditions, the first one being that the user has an approximate understanding of the dominant relations between design variables and target parameters, and that simulations or measurements can be executed to obtain information of what parameters the current design results in. We will refer to this approach as the {\bf A}pproximate {\bf N}onlinear {\bf Mo}del-{\bf b}ased ({\bf ANMod}) optimization method. 

We provide the full implementation of the optimizer as an open-source Python package QDesignOptimizer \cite{qdesignoptimizer_git}, which offers a flexible and modular structure for efficient subsystem optimizations. Simulation routines and analytical models are dynamically compiled on the basis of user-defined settings. To demonstrate the practical applicability of our package, we include several examples for single and two-qubit chip design optimizations. 

\section{Physics-guided, multi-parameter optimization}
\begin{figure}
    \centering
    \includegraphics[width=1\linewidth]{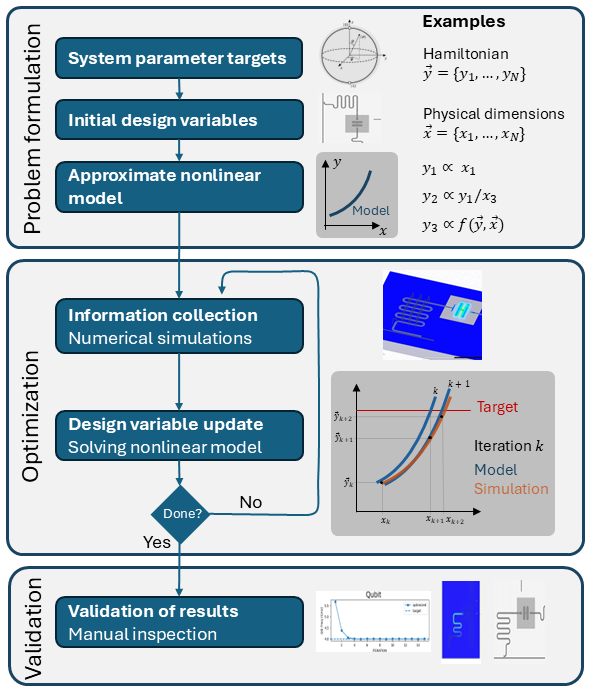}
    
    \caption{Design flow of the QDesignOptimizer consisting of problem formulation, optimization and validation. In the problem formulation, the user specifies the system target parameters and initial values of the design variables as well as the nonlinear relationship between them. The optimization starts by a numerical simulation using the current design variables $x_{k}$ to study which parameter values $y_{k}$ they correspond to. The design variables are then updated by solving the nonlinear model  evaluated with the currently available information such that $\overrightarrow f( \overrightarrow{y_{k}},  \overrightarrow{x})=\overrightarrow y^{target}$, resulting in the updated values $x_{k+1}$. The numerical simulation runs again, to predict $y_{k+1}$ providing a point closer to the target at which the model can be evaluated at. 
    Finally, the user manually inspects the convergence, mode shapes and resulting design for any undesired properties.}
    \label{fig:Fig1_flow}
\end{figure}

The workflow of our ANMod-optimization framework, QDesignOptimizer, depicted in Fig.~\ref{fig:Fig1_flow}, consists of the problem formation step, the actual optimization step, and the validation step. 
In the problem formulation step, the user defines (i) the target parameters, (ii) an initial guess for values of the design variables, and (iii) the approximate physical relations as a multi-parameter nonlinear model between the parameters and design variables.
In the optimization step, the set of design variables is iteratively updated based on the solution of the approximate nonlinear model. In every iteration, the current set of parameters is computed from simulations and further analysis. 
In the validation step, the result is finally manually inspected for convergence and general feasibility. 
\\In the following subsections, we will introduce the general mathematical framework.

\subsection{Problem formulation} \label{sec:problem_formulation}
The problem formulation begins with the definition of $N$ design variables $ \overrightarrow{x}=\{x_1, ..., x_N\}$, which are varied to reach the targets of the $N$ parameters $ \overrightarrow{y}=\{y_1, ..., y_N\}$. These $N$ definitions of design variables and parameters are linked through $N$ approximate nonlinear models, 
\begin{equation}
     y_i\propto f_i\left(\underset{\textrm{excl.} y_i}{ \overrightarrow{y}},  \overrightarrow{x}\right) 
     \label{propto}
\end{equation}
which specify the proportionality relation between the parameter and the design variables as well as (other) parameters.
 
The framework enables the user to easily change both the physical relationships as well as the dimensionality of the problem by selecting which parameters to optimize. In complex systems, the model refinement could potentially be guided by an empirical numerical study of the system's behavior. Finally, as an initial guess for the optimization, the user has to provide sensible values for the design variables $\overrightarrow{x}$. The design variables can be constrained during the optimization to enhance stability. 

\subsection{Information collection - simulate design variables} \label{sec:information_collection}
To obtain the parameter values $\overrightarrow{y}^{k}$ for the current set of design variables $\overrightarrow{x}^{k}$, the optimization framework needs to run some detailed simulations and further analysis (or potentially measure physical devices). In the QDesignOptimizer, detailed electro-magnetic eigenmode or capacitance simulations are run in Ansys HFSS, and followed by an energy-participation ratio (EPR) analysis. 
The parameters $\overrightarrow{y}^{k}$, in the $k$th iteration, are then extracted from the simulation and analysis results. 

\subsection{Update design variables} \label{sec:parameter_update}
Here, we present the core idea of the ANMod-optimization method. To reach the parameter targets through the optimization, we aim to efficiently update the design variables from the ones used in the simulation $\overrightarrow{x}^{k}$ in iteration $k$ to $\overrightarrow{x}^{k+1}$, such that $\overrightarrow{y}^{k+1}$ gets close to the target value $\overrightarrow{y}^{target}$. Given the proportionality between $y_i$ and $f_i$, the parameter corresponding to the updated design variable can be estimated by
\begin{equation}
   \tilde y_i^{k+1} = y_i^{k} \frac{f_i(\overrightarrow{\tilde y}^{k+1},\overrightarrow{x}^{k+1})}{f_i(\overrightarrow{y}^k,\overrightarrow{x}^k)}.
   \label{Qk1}
\end{equation}
Here, we distinguish the quantity $ \overrightarrow{\tilde y}$ (with $\tilde {}\ $) estimated by the approximate model (Eq.\ref{propto}) and the simulated quantity $ \overrightarrow{y}$, which typically is more accurate. 
Hence, the more accurate the nonlinear model set in Eq.\ref{propto}, the smaller is the discrepancy $|| \overrightarrow{y}- \overrightarrow{\tilde y}||$ and the faster the optimization converges. 
To obtain the updated design variables $\overrightarrow{x}^{k+1}$, we minimize the cost function
  \begin{equation}
 \label{cost}
C = \sum_{i=1}^N\left|\frac{\tilde y_i^{k+1}}{y_i^{target}} - 1\right|^2,
 \end{equation}
 by finding the optimal $\overrightarrow{x}^{k+1}$. If the problem is correctly formulated, the minimization will reach 
 \begin{equation}
     \tilde y_i^{k+1} = y_i^{target}
 \end{equation} 
 for all $N$ targets in the optimization. (Note however that user constraints on the design variables might result in a cost $C>0$). Hence, the closed expression minimized by the ANMod-method in each iteration is 
  \begin{equation}
 \label{cost}
C = \sum_{i=1}^N\left|\frac{y_i^{k}}{y_i^{target}} \frac{f_i(\overrightarrow{\tilde y}^{target},\overrightarrow{x})}{f_i(\overrightarrow{y}^k,\overrightarrow{x}^k)} - 1\right|^2
 \end{equation}
which only depends on the three inputs: (i) the values of the parameters in the previous step $\overrightarrow{y}^k$, (ii) the target parameter values $\overrightarrow{y}^{target}$, and (iii) the design variables in the previous step $\overrightarrow{x}^k$. The optimizer then finds the variables $\overrightarrow{x}$ which minimize the cost $C$. The parameter update can be low-pass filtered with the adjustment-rate setting for more stable optimization.

Parameters can be flexibly included or excluded from the optimization without the need to rewrite other relations that depend on them. To support this, the QDesignOptimizer assumes that parameters without specified optimization targets remain unaffected by changes in the design variables, i.e., $\tilde y_i^{k+1} = y_i^{k}$ for $i > N$.

\subsection{Design validation}
The final step involves a manual inspection of the optimization results to ensure the feasibility of the optimized design. This includes verifying the convergence of the information-collecting simulations as well as the overall optimization towards the target parameters, inspecting the spatial mode shapes to confirm correct localization, and evaluating the resulting design to ensure that it meets any requirements that were not explicitly specified in the optimization targets. Note that the convergence of the optimizer towards the target parameters is not guaranteed in general (see Sec.~\ref{sec:model_init_conditions}).

\subsection{Comparison to other multi-parameter optimizers}
For our physics–guided ANMod-optimizer, the time it takes for the algorithm to convergence is limited by the time required to execute the simulations needed to evaluate the parameters. Consequently, precise design variable updates are essential to reduce the number of required simulation iterations for the nonlinear $N$–dimensional minimization problem. In superconducting circuit design, physical models often provide reliable approximations for quantities such as frequencies and coupling strengths, which we leverage to accurately update the design variables. However, even a crude nonlinear model that correctly captures the qualitative trend of the target response is often enough to obtain fast convergence. As for the minimization, the optimizer will separate and solve independent variables whenever possible, thereby reducing the effective dimensionality of the search space.

Our QDesignOptimizer is based on the ANMod-update strategy; in that, it markedly differs from general-purpose optimizers, which instead rely on derivative-free or gradient-based methods. These methods require function or gradient computation through either targeted simulations triggered by the optimizer, or pre-computed libraries of design evaluations obtained from parameter sweeps. 
Both approaches typically require the evaluation of significantly more simulations compared to our ANMod-method, in which the gradient information is directly incorporated into the optimizer through the underlying, nonlinear physical model.
At the same time, it is worth pointing out that all these methods, including the one presented in this work, may converge to a local rather than a global optimum.

\section{Application in circuit quantum electrodynamics}

The QDesignOptimizer package using the ANMod-method can be applied to a broad range of systems in circuit quantum electrodynamics, such as superconducting qubits, resonators, couplers, and combinations of those. 
In this section, we demonstrate its usefulness and limitations on two representative examples.
We begin with the fundamental building block of superconducting quantum processors -- a planar qubit–resonator system -- to introduce the central ideas of eigenmode–analysis–based optimization. 
We then examine how nonlinear models and initial conditions can influence the optimization outcomes. 
To illustrate the scalability of the method to larger design spaces, we apply the optimizer to a more complex qubit–coupler–qubit device with five modes. 
Finally, we discuss how modular optimization and capacitance–simulation-based optimization extend the applicability of the QDesignOptimizer. 
For a detailed code-level tutorial, we refer the reader to the online documentation of our QDesignOptimizer python package \cite{qdesignoptimizer_git}.

\subsection{Eigenmode–Analysis–Based Optimization}

For the optimization of the single qubit-resonator system, the problem formulations follows the ANMod-method outlined in Sec.~\ref{sec:problem_formulation}. First, a sketch of the system is designed in Qiskit-Metal. Then, the optimizer takes the information summarized in Table~\ref{tab:quantity_table}. In the given case, the five target parameters are the resonator frequency, resonator decay rate, qubit frequency, qubit anharmonicity and resonator-qubit dispersive shift. The same number of design variables are introduced, each with an initial value that can be varied to reach the target. Note that depending on the choice of design variables, one design variable might affect several parameters, e.g. $w_{qb}$ is primarily used to set the anharmonicity, but it simultaneously affects the qubit frequency (see also App.~\ref{app:recommendation}). Finally, an approximate physical relation is provided for each parameter, describing its relationship to the design variables and other system parameters. For instance, the resonator frequency is parameterized by its physical length.

\begin{table*}
    \caption{Summary of the problem formulation for the single qubit-resonator system. The design variables are resonator length $l_{res}$, qubit Josephson junction inductance $L_{qb}$, qubit width $w_{qb}$, resonator-qubit coupling width $w_{res-qb}$ and resonator to transmission line coupling length $l_{res-tl}$. To further detail the model of the dispersive shift, we directly input the parameters anharmonicity $\alpha$ and difference frequency $\Delta =f_{qb}-f_{res}$.}
    \label{tab:quantity_table}
    \begin{tabular}{|c|c|c|c|c|c|}
    \hline
    \textbf{Parameter} & \textbf{Symbol} & \textbf{Target} & \textbf{Proportional to} & \textbf{Design variable}  & \textbf{Initial value} \\
    \hline
    \hline
    Resonator frequency & \( f_{res} \) & 6 GHz & \( 1 / l_{res} \) &  \( l_{res} \) &  7500 \(\mu\)m  \\
    \hline
    Qubit frequency & \( f_{qb} \)  & 4 GHz & \( 1 / \sqrt{L_{qb} \cdot w_{qb}} \) &  \( L_{qb}\)  &  12.1 nH \\
    \hline
    Anharmonicity & \( \alpha \) & 200 MHz & \( 1 / w_{qb} \) &  \( w_{qb}\) & 400 \(\mu\)m \\
    \hline
    Dispersive shift & \( \chi_{qb-res} \) & 1 MHz & \( (w_{res-qb}/w_{qb})^2 \cdot \alpha / [\Delta (\Delta -\alpha)] \)  &  \( w_{res-qb}\)  & 100 \(\mu\)m\\
    \hline
    Resonator decay rate & \( \kappa_{res} \) & 1 MHz & \( l_{res-tl} \) & \( l_{res-tl} \) &  400 \(\mu\)m\\
    \hline    
    \end{tabular}
\end{table*}

The subsequent optimization follows the iterative approach of information collection and design variable updates described in Sec.~\ref{sec:information_collection} and Sec.~\ref{sec:parameter_update}, respectively. First, the design is simulated using Ansys HFSS's eigenmode solver, followed by energy participation ratio (EPR) analysis through pyEPR, both accessed directly via the Qiskit-Metal integration. While the present example uses the integrated HFSS/pyEPR toolchain, the QDesignOptimizer can be extended to other simulation backends to support diverse academic and industrial workflows.

\begin{figure*}
    \centering    \includegraphics{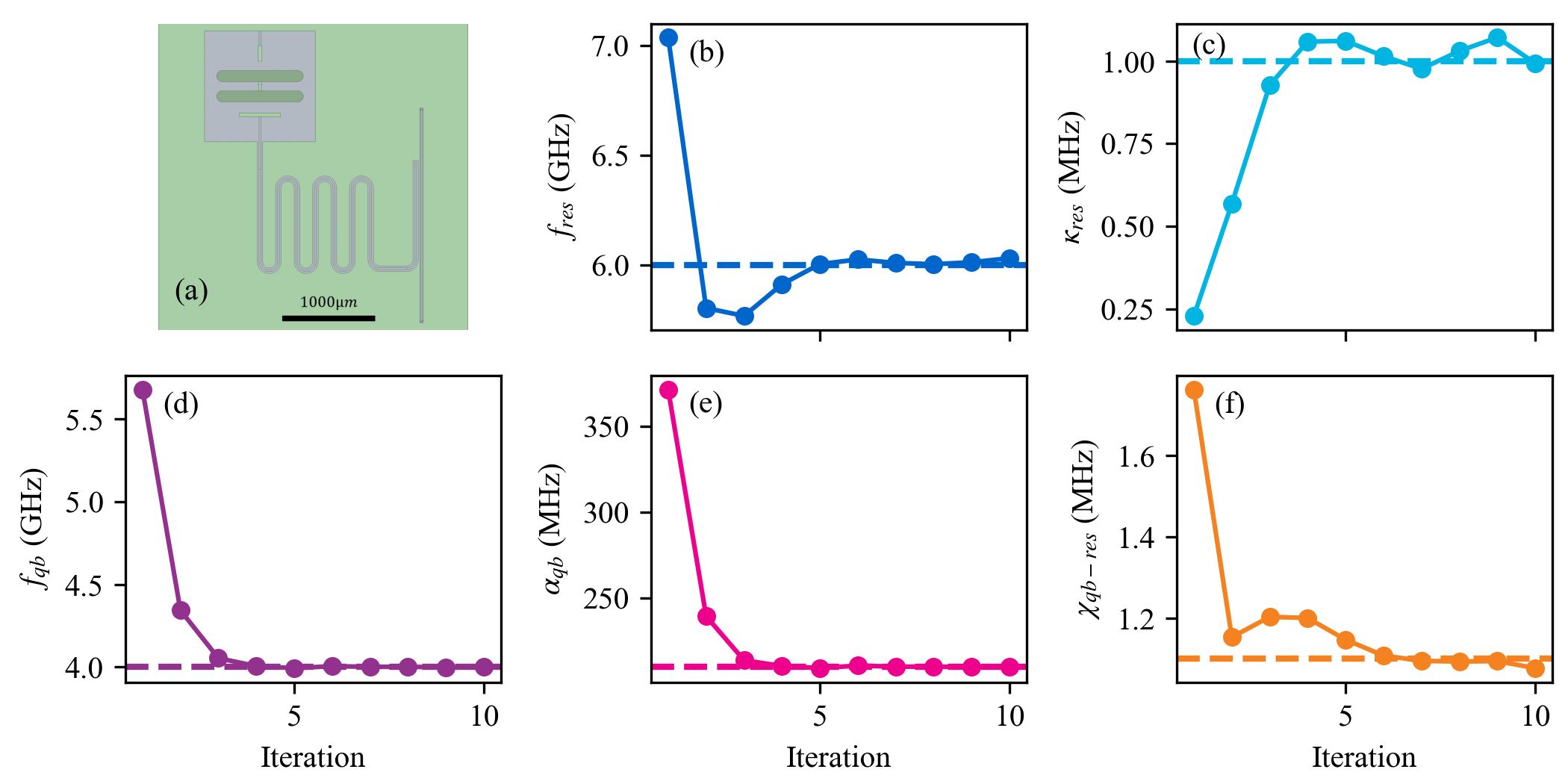}    
    \caption{Eigenmode-analysis-based optimization of a qubit-resonator-waveguide system
    with the problem formulation described in Tab.~\ref{tab:quantity_table}. (a) Screen capture of the simulated structure in Ansys HFSS. (b-f) Optimization results versus iterations for the five target parameters. The dashed lines indicate the parameter targets. In each iteration, HFSS performed 8 mesh refinement passes. }
    \label{fig:fig2_OPT_single_qubit_resonator}
\end{figure*}

The results for the optimization of the whole subsystem described in Tab.~\ref{tab:quantity_table} and sketched in Fig.~\ref{fig:fig2_OPT_single_qubit_resonator}a are shown in the Figures~\ref{fig:fig2_OPT_single_qubit_resonator}b-f per iteration. All five target parameters converge within a few iterations.
The resonator decay rate, $\kappa$ exhibits fluctuations even after several iterations [Fig.~\ref{fig:fig2_OPT_single_qubit_resonator}(c)]. We ascribe these to the inaccuracy of the $\kappa$ estimate from the imaginary part of the resonator mode analysis, which, in turn, increases the error in the design variable update (see App.~\ref{sec:numer_of_passes}). The observed fluctuations of $\kappa$ also indirectly affect the other targets, especially the resonator frequency.

\subsection{Effect of Nonlinear Models and Initial Conditions} \label{sec:model_init_conditions}
Following the eigenmode–analysis–based optimization example in the previous section, a natural question arises: how accurate must the nonlinear model be to ensure efficient and rapid convergence? To address this, we compare several optimization runs with identical initial conditions but different nonlinear models for the resonator–qubit dispersive shift $\chi$.

\begin{figure}
    \centering    \includegraphics{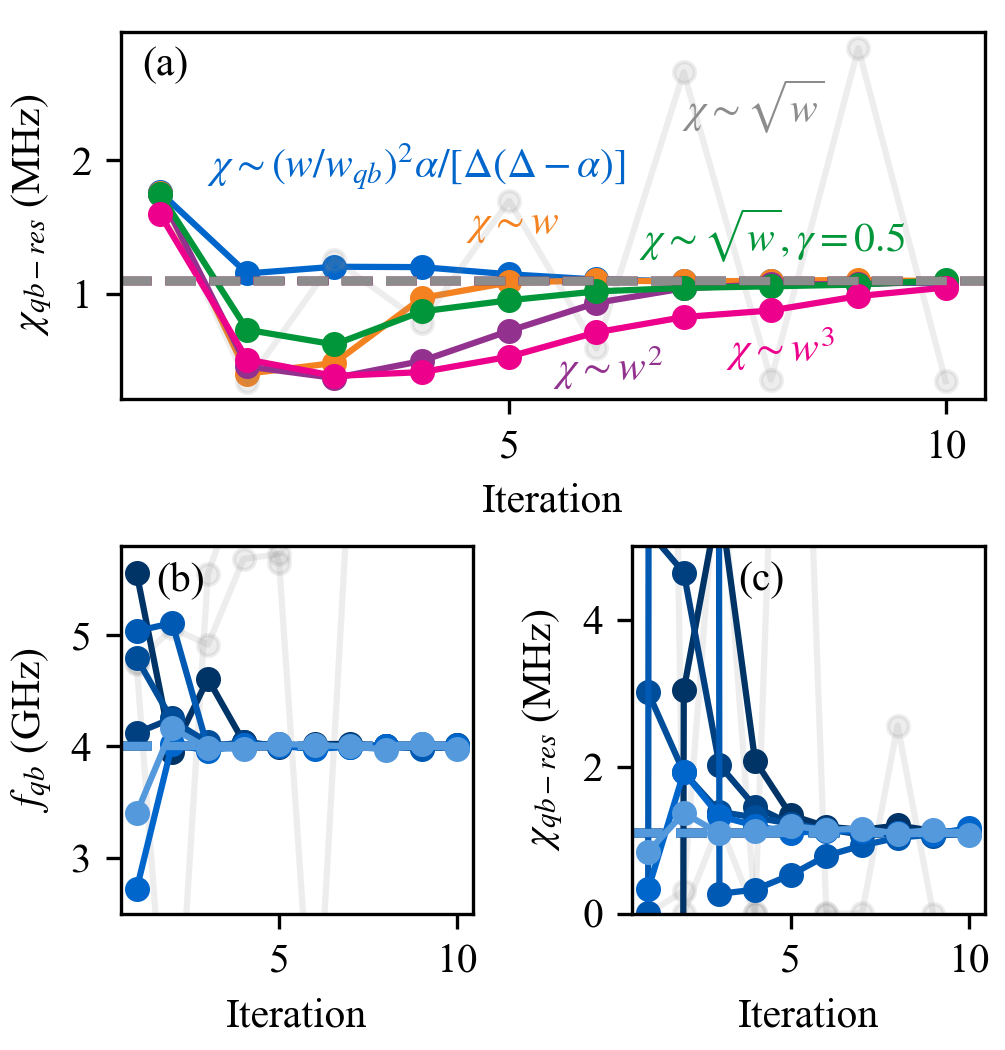}
    \caption{Effect of different nonlinear models and different initial conditions on optimization results obtained from an eigenmode-analysis-based optimization of a resonator-qubit system coupled to a feedline as in Fig.~\ref{fig:fig2_OPT_single_qubit_resonator}. (a) Optimization result for the resonator-qubit $\chi$ for five different nonlinear models. All models leads to convergence except $\chi\propto \sqrt{w}$, which diverges since this model heavily underestimates the sensitivity on $w$, here denoting $w_{res-qb}$. However, reducing the adjustment rate to 50\%, i.e. low-pass filtering the updates, restores convergence even for this model. For each iteration, 8 simulation passes were used. (b,c) Optimization results for (b) the qubit frequency and (c) resonator-qubit dispersive shift $\chi$ for ten different initial conditions using the full $\chi$ expression as input to the nonlinear model. Each set of initial conditions is encoded with the same shade of blue for the eight successful optimization runs. The remaining two failed optimizations we label in gray. The dashed lines indicate the parameter target. In each iteration, HFSS performed 4 mesh refinement passes. }
    \label{fig:fig3_single_qubit_sample}
\end{figure}

Strikingly, both a very precise model and a variety of less accurate ones converge in a few iterations as shown in Fig.~\ref{fig:fig3_single_qubit_sample}a. As seen, the most accurate expression used here $\chi=(w/w_{qb})^2\alpha/(\Delta(\Delta-\alpha))$ \cite{Blais_2021}, accurately accounts for the simultaneous optimization of the mode frequencies (affecting $\Delta$ and therefore $\chi$), leading to the fastest convergence. In contrast, none of the other models captures this phenomenon and therefore overshoots considerably after the first design variable update. After two updates, the other parameters are close to their target, which allows simpler models to be effective from here on as well. 
The model $\chi\propto w$ proofs itself most effective under the circumstances and rapidly converges $\chi$ from there on. 
The model $\chi\propto w^2$ assumes a too sensitive dependence on $w$ and therefore suggests a too small update step leading to slower convergence, an effect even stronger for $\chi\propto w^3$. 
Analogously, the model $\chi\propto \sqrt{w}$ assumes a too insensitive dependence on $w$ and therefore suggests a too large update step which overshoots to the extent that the optimization diverges. 
However, combining the too inaccurate model $\chi\propto \sqrt{w}$ with low pass filtering via the adjustment rate $\gamma$ of the optimization, convergence can be restored. 
Furthermore, by studying how convergence is reached, the user can get a better understanding of a more appropriate model. To summarize, in general, as long as the nonlinear model captures the correct qualitative trends and the step size is feasible, the optimizer will converge. 

A second question concerns the robustness of the optimization with respect to different initial conditions, since convergence is generally not guaranteed for nonlinear systems. 
To investigate this, we perform multiple optimizations of the same resonator–qubit system defined in Tab.~\ref{tab:quantity_table}, but with varying initial values of the design variables. The choice of initial conditions was random, but constrained to the geometric limitations allowed by the design, i.e. the box of the split-transmon and the separation between the launchpads, see App.~\ref{app:design_variable_constraints}. 
Figures~\ref{fig:fig3_single_qubit_sample}b-c show the convergence trajectories for two parameters of our example design under ten different choices of initial design variables. The same set of initial conditions are encoded with the same shade of blue in the two sub-figures. In eight out of then tested cases, the optimizer quickly reaches the target parameter values despite the wide range of initial conditions, demonstrating strong robustness against variations in the initial conditions.
However, in two out of ten at random tested cases, the optimization did not converge over ten iterations or got aborted by the optimizer.    

To avoid the most common failure modes, certain constraints must be respected. 
First, geometric overlaps during the automatic routing of structures must be prevented; this can be enforced by specifying design variable bounds to avoid physically infeasible layouts. 
Second, identical ordering of the $M$ modes specified in the optimizer setup and obtained in the HFSS simulation should be maintained throughout the optimization to avoid incorrect parameter update due to misassigned modes. This problem can be identified from the field distribution in HFSS. To avoid unintended mode crossings, we generally advise to start the optimization with well separated mode frequencies. 
Third, if the underlying physical relationships in the nonlinear model are fundamentally incorrect, convergence cannot be expected, as the optimizer’s predictions will consistently point away from or overshoot beyond the true optimum. In the case of non-monotonic nonlinear models or of counter-acting design variables, the optimization might get stuck in a local minimum.

\subsection{Optimization of a Two-Qubit Device}
\label{sec:2qubitexample}
\begin{figure}
    \centering
    \includegraphics{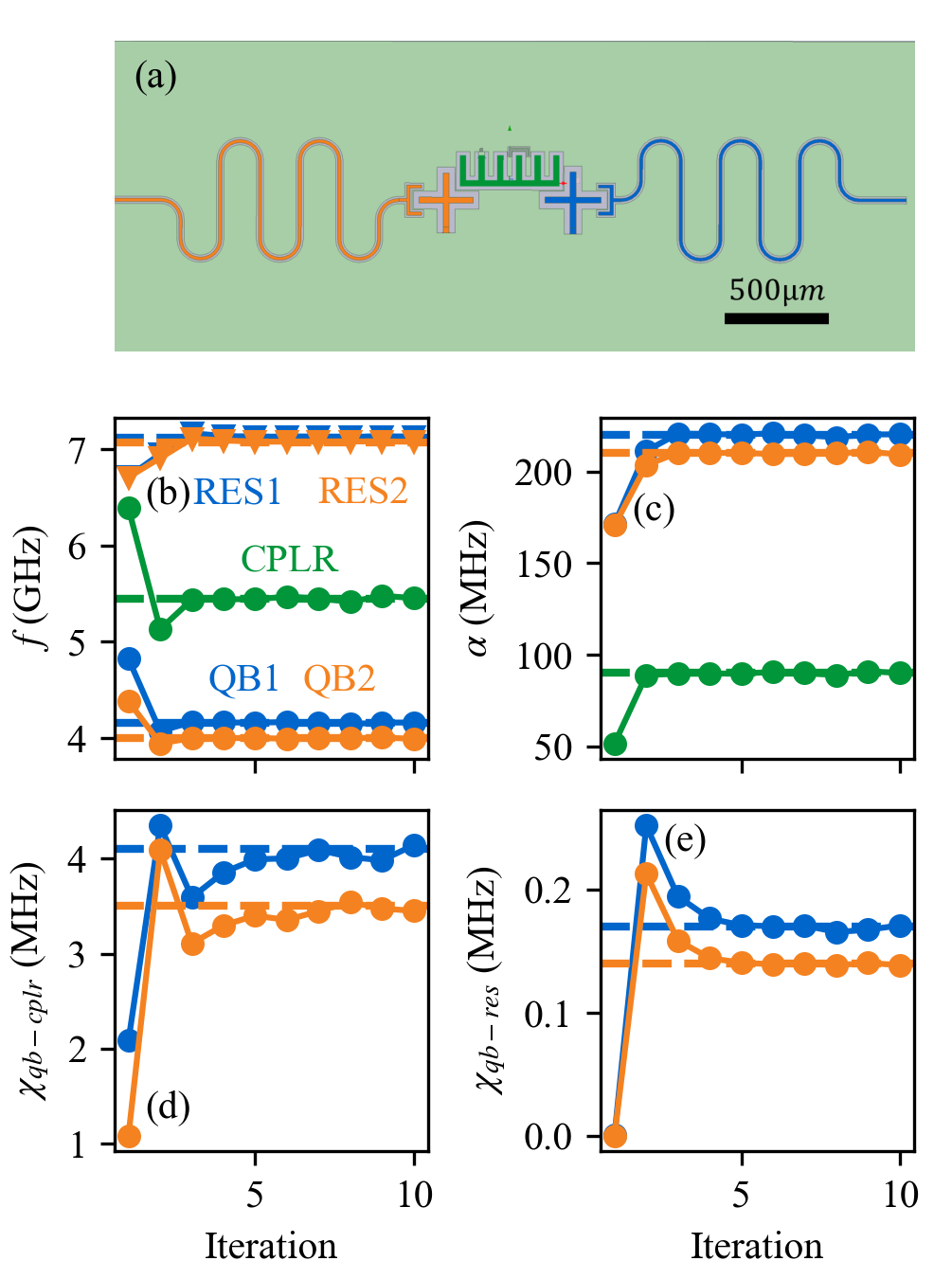}
    \caption{Eigenmode-analysis-based optimization of a system composed of two qubits coupled via a tunable coupler and two readout resonators, as studied in Ref.\cite{Sung_2021}. The problem formulation is described in App.~\ref{app:opt_setup_twp_qb_example} consisting of simultaneous optimization of 12 parameters for 5 modes. }
    \label{fig:Fig4_OPT_two_qubit_resonator}
\end{figure}

To demonstrate the effectiveness of the QDesignOptimizer in tackling larger, multi-mode systems, we apply it to a two-qubit quantum processor as in Ref.~\cite{Sung_2021}. The system comprised of two X-mon qubits with nonlinear coupler and attached readout resonators includes five modes and 12 target parameters. The detailed problem formulation is described in App.~\ref{app:opt_setup_twp_qb_example}. Given this setup, the optimization quickly converges for all parameters as shown in the convergence trajectories versus iterations in Fig.~\ref{fig:Fig4_OPT_two_qubit_resonator}. 
As another tool tip, once a suitable problem formulation has been found, it is often possible to re-run the optimization with new parameter targets without further adjustments, given that the mode order in frequency is preserved. This significantly reduces the effort required to change the target parameters.
Note that it took one of our QDesignOptimizer developers less than two work days to create the design, setup the problem formulation and reach acceptable convergence. Hence, we believe that the QDesignOptimizer package will prove itself effective when tuning up qubit circuits such as in Refs.~\cite{Sheldon_2016, McKay_2016, Noguchi_2020, Moskalenko2022, Warren2023}. An improvement to the example shown here could be the addition of a direct capacitive coupling between the two qubits to cancel their cross ZZ interaction as yet another target parameter\cite{Fors2024}. 

\subsection{Modular Optimization}
Scaling to larger processor designs with multiple qubits and couplers quickly becomes infeasible to simulate in full on a desktop computer. To address this, the QDesignOptimizer workflow supports a modular optimization strategy: First, the chip design can be partitioned into subsystems by selecting relevant Qiskit-Metal components, ports, and modes. Then, the subsystems are iteratively optimized for overlapping sets of coupled modes. Since circuit layouts vary widely, there is no general recipe for the decomposition into subsystems for arbitrary design configurations. 
In practice, strongly interacting subsystems are optimized in isolation under the assumption that the influence of excluded structures is negligible. If computationally feasible, the resulting design can be validated through a single, more expensive simulation of the larger system. 
For instance, instead of simulating a complete multi-qubit device with couplers and readout resonators, one may first optimize smaller building blocks—such as the qubit–resonator system in Fig.~\ref{fig:fig2_OPT_single_qubit_resonator}a—before addressing larger subsystems like the qubit–coupler–qubit configuration like in Fig.~\ref{fig:Fig4_OPT_two_qubit_resonator}a. 

\subsection{Capacitance–Simulations–Based Optimization}
In addition to eigenmode–analysis–based optimization, which supports targets for mode frequencies, inter- and intra-mode nonlinearities, and mode decay rates, the QDesignOptimizer supports optimization workflows based on capacitance simulations. This feature enables the direct targets of capacitances between islands, and derived targets, such as the resonator–feedline coupling rate $\kappa$ or the $T_1$ limit due to energy decay from a transmon qubit into a charge line. We present an example for the capacitance-simulation-based optimization in App.~\ref{app:capacitance_target}. Already at present, users may define their own custom targets, leveraging results from both eigenmode and capacitance simulations in the same optimization cycle. 

\section{Conclusion}
We have presented a physics–guided, multi–variable optimization framework, the QDesignOptimizer, for the automated design of superconducting quantum devices, and demonstrated its use for the optimization of qubit devices in circuit quantum electrodynamics. The framework integrates high–accuracy electromagnetic simulations with user–defined nonlinear models, enabling multi–parameter optimization featuring remarkably stable and fast convergence due to the introduced ANMod-method. Our current implementation is tightly integrated with the Qiskit-Metal and pyEPR toolchain.  

The optimizer does not reduce the raw simulation time of the high-fidelity electromagnetic simulations compared to manual design variable tuning. However, it substantially improves research efficiency by enabling unattended optimization runs, provided the same physical knowledge is supplied as input. The increase in efficiency makes late-stage design modifications more feasible and time-effective, and it reduces the effort associated with re-optimizing complex superconducting quantum devices. 
In future developments, we plan to extend the framework to support scattering–parameter simulations to efficiently simulate the resonator-to-transmission line coupling. 
Note again, that the underlying ANMod-optimization method is general and can be applied to other simulation types and nonlinear systems to solve various engineering problems.

\section*{Code availability} The code for the QDesignOptimizer is available at \cite{qdesignoptimizer_git}.

\section*{Acknowledgment}
We thank the other members of the 202Q-lab for valuable discussions and testing. This work was supported by the Knut and Alice Wallenberg Foundation through the Wallenberg Centre for Quantum Technology (WACQT) and by the European Union’s Horizon Europe Framework Programme (EIC Pathfinder Challenge project Veriqub) under Grant Agreement No. 101114899. S.G. acknowledges financial support from the European Research Council (Grant No. 101041744 ESQuAT).
External interest disclosure: S.G. is a co-founder and equity holder in Sweden Quantum AB.

\appendix
\renewcommand{\thefigure}{S\arabic{figure}}
\setcounter{figure}{0}

\section{Capacitance–Simulations–Based Optimization} \label{app:capacitance_target}

The QDesignOptimizer also supports capacitance targets or targets derived thereof. Capacitance optimizations can be run separately or jointly with the eigenmode simulations. 
Two already built-in derived target parameters are the $T_1$ limit due to energy decay from a transmon qubit into a charge line, and the resonator–feedline coupling rate $\kappa$. In future work additional derived targets, including Purcell-limited decay and dephasing due to thermal photons in the readout resonator could be added. Already at present, users may define their own custom targets, leveraging results from both eigenmode and capacitance simulations.

\begin{figure}
    \centering
    \includegraphics{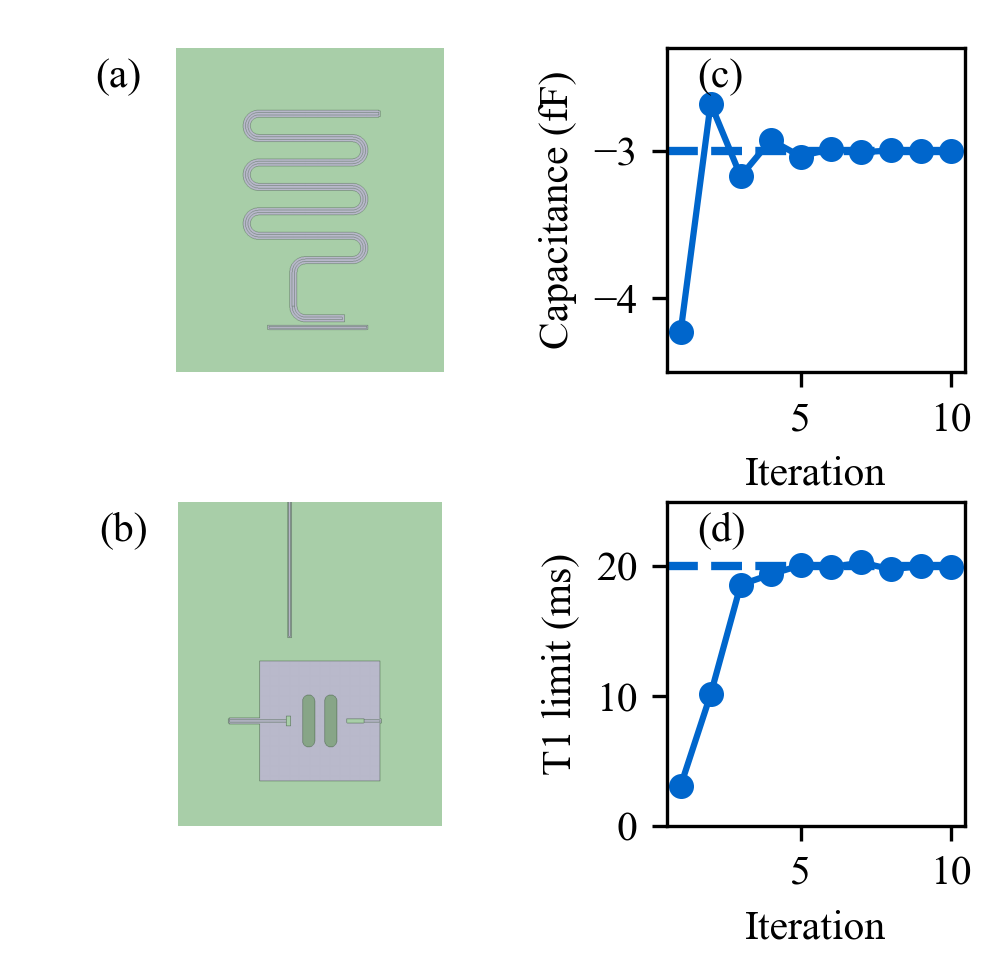}
    \caption{Capacitance-simulation-based optimizations. (a,c) Screen capture and optimization result for the coupling capacitance between a coplanar waveguide resonator to a finite-length feedline. (b,d) Screen capture and optimization result for the T1 limit of the split-transmon decaying into the neighboring charge line at the top in the presence of other coupling ports to the left and right side. The T1 limit is derived from the full capacitance matrix of the circuit. Per iteration 10 simulation passes were used.}
    \label{fig:FigS1_capacitance_decay}
\end{figure}

Figure~\ref{fig:FigS1_capacitance_decay}a shows an example setup consisting of a coplanar waveguide (CPW) resonator capacitively coupled to a finite-length CPW transmission line. The optimization goal is to achieve a specified target capacitance between the two structures. As the physical model, we assume an inverse square-root dependence of the capacitance on the coupling distance, which serves as the design variable in this case. The resulting optimization, shown in Fig.~\ref{fig:FigS1_capacitance_decay}c, reaches the target value in approximately five iterations for the given initial conditions.

In a second example, we consider a split-transmon qubit with a nearby charge line, as illustrated in Fig.~\ref{fig:FigS1_capacitance_decay}b. Here, the objective is to optimize the $T_1$ limit set by energy decay into the charge line, using the position of the charge-line tip as the design variable. The nonlinear model assumes a cubic dependence of the target parameter, the $T_1$ limit, on the relative position of the charge line. Notably, defining design variables in terms of relative coordinates or distances, rather than absolute positions, often improves optimization robustness, as it reduces sensitivity to global coordinate shifts or layout changes and greatly reduces the complexity of the physical relationship. 
We obtain the $T_1$ limit from the evaluation of a classical model for a single mode decay into a single decay channel taking the capacitance matrix, the mode frequency and the line impedance as input \cite{Pechal2016}.
The optimization result, shown in Fig.~\ref{fig:FigS1_capacitance_decay}d, converges to the target within roughly five iterations, demonstrating the efficiency of capacitance-based optimization workflows for both direct and derived target quantities.

\section{Effects of simulation setup on optimization} \label{sec:numer_of_passes}
\begin{figure}
    \centering
    \includegraphics{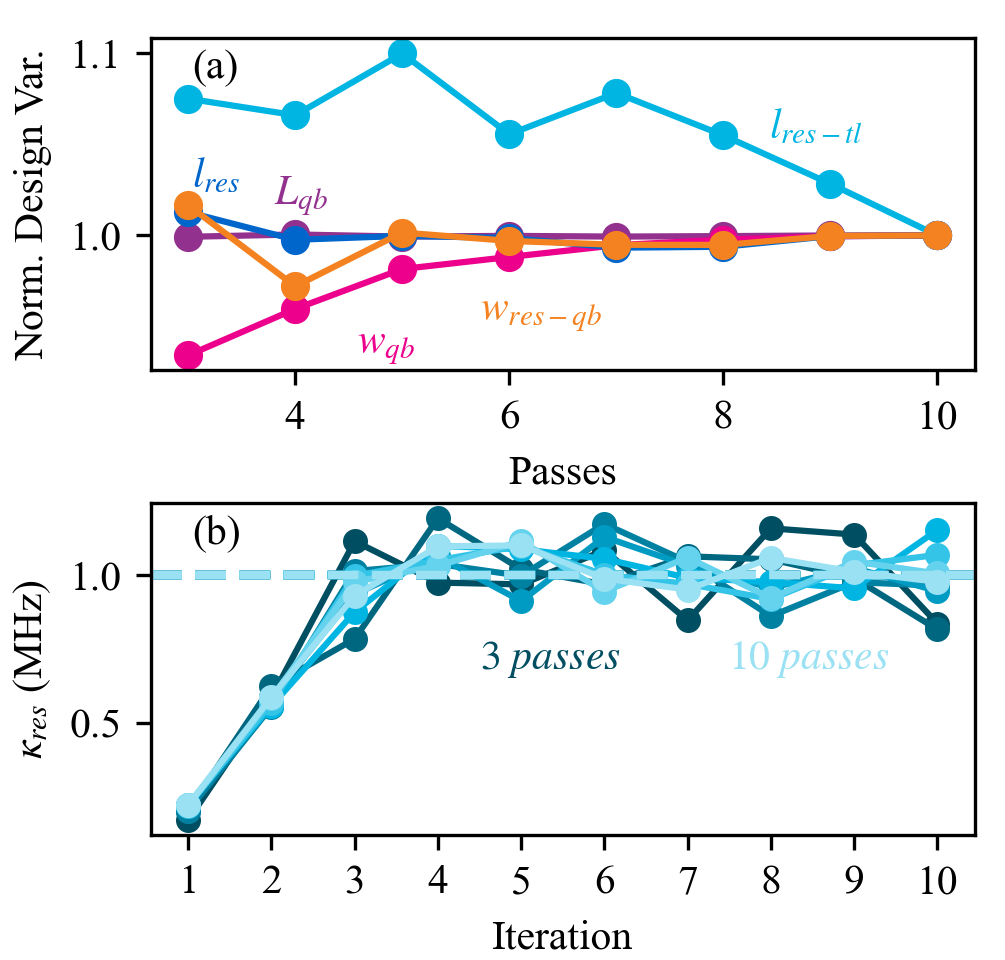}
    \caption{Eigenmode-analysis-based optimization of a resonator-qubit system coupled to a feedline for different numbers of passes. (a) Relative difference of the five design variables obtained after ten iterations of optimization with different numbers of passes normalized to the results obtained from the largest number of passes. The colors of the design variables matche the colors of the optimization targets in Fig.~\ref{fig:fig2_OPT_single_qubit_resonator}. (b) Optimization results versus iterations for the resonator $\kappa$ for different numbers of passes used during the simulation. Each number of passes between 4 and 10 is encoded in a different shade of lightblue. }
    \label{fig:FigS2_OPT_single_qubit_resonator_passes}
\end{figure}

To understand the effect of the accuracy of the parameter data obtained from the simulations on the convergence of the optimization, we perform independent optimizations with 3 to 10 mesh refinement passes in HFSS and compare the final design variables obtained after ten iterations. 
The resulting difference between design variables is shown in Figure~\ref{fig:FigS2_OPT_single_qubit_resonator_passes}a. 
While the resonator length and the qubit’s Josephson inductance remain relatively stable across different pass numbers, the qubit pad length and the resonator–qubit coupling segment length exhibit variations of approximately 3\%. In contrast, the resonator–feedline coupling segment length exhibits variations of up to 10\% relative to the most accurately simulated values obtained with the largest number of passes for this set of initial conditions.
From this graph, it becomes apparent that all design variables require eight or more passes to obtain accurate optimization results in this particular simulation setup. Note that the exact number of required passes depends on the feature sizes, the initial mesh, and the mesh refinement rate. Thus, there is a trade-off: a higher number of passes improves accuracy, but also increases overall optimization time.
Especially the resonator-feedline coupling segment length (lightblue) shows a non-monotonic trend with increasing number of passes, which is primarily due to the inherent limitations in extracting $\kappa$ from HFSS eigenmode simulations via the imaginary part of the resonator eigenmode. HFSS's adaptive meshing per simulation pass prioritizes regions of high electric field rather than areas requiring accuracy for coupling parameter extraction, such as the coupling segment geometry. Hence, a high mesh refinement yielding an accurate $\kappa$ estimate is only reached at a high number of passes.  
As a result of this meshing problem, $\kappa$ exhibits the largest fluctuations, especially for a few number of passes, compared to all other parameters, directly affecting the stability of its corresponding design variable during optimization. The fluctuations of $\kappa$ are visible in  Figure~\ref{fig:FigS2_OPT_single_qubit_resonator_passes}b, which shows the $\kappa$ estimates throughout the optimizations for different number of passes encoded in different shades of lightblue.
To partially mitigate the problem of insufficient mesh refinement in the resonator-feedline coupling region, we manually specify a finer, initial meshing map locally around the coupling segment in addition to the coarse default initial mesh assigned by HFSS in all simulations shown in this paper. With this additional fine initial mesh, a ten-iteration optimization with eight simulations passes each requires approximately 280 minutes on one of our office workstation. 

\subsection{Design Variable Constraints for Initial Conditions used in single qubit example} \label{app:design_variable_constraints}

The ten different sets of initial conditions used for Fig~\ref{fig:fig3_single_qubit_sample}b-d were uniformly sampled within the boundaries listed in Table~\ref{tab:design_constraints}. These limits reflect to the maximal geometric extend allowed by the allocation of feedline ports and qubit placement.

\begin{table}[h]
    \centering
    \caption{Design variable constraints used to sample the initial conditions seeding the optimizations shown in Fig~\ref{fig:fig3_single_qubit_sample}b-d.}
    \label{tab:design_constraints}
    \begin{tabular}{|l|c|c|}
    \hline
    \textbf{Design Variable} & \textbf{Min} & \textbf{Max} \\
    \hline \hline
    Qubit pad Width & 100~$\mu$m & 1100~$\mu$m \\
    Qubit Josephson Inductance & 5~nH & 25~nH \\
    Resonator Length & 4000~$\mu$m & 12000~$\mu$m \\
    Qubit-Resonator Coupling Length & 100~$\mu$m & 1100~$\mu$m \\
    Resonator- Feedline Coupling Length & 100~$\mu$m & 1400~$\mu$m \\
    \hline
    \end{tabular}
\end{table}

\section{Optimization setup used in the two-qubit example} \label{app:opt_setup_twp_qb_example}
The problem formulation for the two-qubit-coupler example discussed in Sec.~\ref{sec:2qubitexample} is summarized in Tab.~\ref{tab:quantity_table_qb_qb}. The setup is notably similar to the problem formulation of the single-qubit-resonator example. The target parameter values were derived from Ref.~\cite{Sung_2021}.
\begin{table*}
    \caption{Summary of the problem formulation for the two qubit quantum processor studied in \cite{Sung_2021}. The qubit-coupler couplings are obtained assuming the ball-park relation $\chi=g^2/\Delta$. The parameters for the qubit-resonator systems are the same as in  Table\ref{tab:quantity_table} with subscripts $i=1,2$, apart from that $w_{res-qb,i}$ is here the length of the horizontal claw segment constrained to $<80\mu$m. Additionally, we have introduced the design variables coupler inductance $L_c$, coupler finger length $w_c$, and diagonal distance of the coupler-qubit metal strip separation $w_{c-qb,i}$.}
    \label{tab:quantity_table_qb_qb}
    \begin{tabular}{|c|c|c|c|c|c|}
    \hline
    \textbf{Parameter} & \textbf{Symbol} & \textbf{Target} & \textbf{Proportional to} & \textbf{Design variable}  & \textbf{Initial value} \\
    \hline
    \hline
    Resonator frequency & \( f_{res,1} \) & 7.12 GHz & \( 1 / l_{res,1} \) &  \( l_{res,1} \) &  4200 \(\mu\)m  \\
    \hline
    Resonator frequency & \( f_{res,2} \) & 7.07 GHz & \( 1 / l_{res,2} \) &  \( l_{res,2} \) &  4200 \(\mu\)m  \\
    \hline
    Qubit frequency & \( f_{qb,1} \)  & 4.16 GHz & \( 1 / \sqrt{L_{qb,1} \cdot w_{qb,1}} \) &  \( L_{qb}\)  &  10 nH \\
    \hline
    Qubit frequency & \( f_{qb,2} \)  & 4.00 GHz & \( 1 / \sqrt{L_{qb,2} \cdot w_{qb,2}} \) &  \( L_{qb,2}\)  &  12 nH \\
    \hline
    Coupler frequency & \( f_{c} \)  & 4.00 GHz & \( 1 / \sqrt{L_{c} \cdot w_{c}} \) &  \( L_{c}\)  &  2.0 nH \\
    \hline
    Anharmonicity & \( \alpha_1 \) & 220 MHz & \( 1 / w_{qb,1} \) &  \( w_{qb,1}\) & 170 \(\mu\)m \\
    \hline
    Anharmonicity & \( \alpha_2 \) & 210 MHz & \( 1 / w_{qb,2} \) &  \( w_{qb,2}\) & 170 \(\mu\)m \\
    \hline
    Anharmonicity & \( \alpha_c \) & 90 MHz & \( 1 / w_{c} \) &  \( w_{c}\) & 250 \(\mu\)m \\
    \hline
    Dispersive shift & \( \chi_1 \) & 0.17 MHz & \( (w_{res-qb,1}/w_{qb,1})^2 \cdot \alpha_1 / [\Delta_1 (\Delta_1 -\alpha_1)] \)  &  \( w_{res-qb,1}\)  & 50 \(\mu\)m\\
    \hline
    Dispersive shift & \( \chi_2 \) & 0.14 MHz & \( (w_{res-qb,2}/w_{qb,2})^2 \cdot \alpha / [\Delta_2 (\Delta_2 -\alpha_2)] \)  &  \( w_{res-qb,2}\)  & 50 \(\mu\)m\\
    \hline
    Coupler-qubit coupling & \( \chi_{c1} \) & 4.1 MHz & \( 1/[w_{c-qb,1}(f_c-f_{qb,1})] \)  &  \( w_{c-qb,1}\)  & $13$ \(\mu\)m\\
    \hline
    Coupler-qubit coupling & \( \chi_{c2} \) & 3.5 MHz & \( 1/[w_{c,qb,2}(f_c-f_{qb,1})] \)  &  \( w_{c,qb,2}\)  & $15$ \(\mu\)m\\
    \hline
  
    \end{tabular}
\end{table*}

\section{Recommendations for efficient optimizations} \label{app:recommendation}

For the setup of efficient optimizations through our framework, we present in the following a few recommendations. 

\subsection{Make design variables independent}
The optimization problem will be much more stable and easier to describe mathematically if the inter-dependence of design variables and parameters are reduced to the level that they can be neglected. Strive to ensure that each design variable has a significant effect on only one target parameter. Note that one design variable can control multiple physical dimensions in a design, which might help in making the design variables independent. 
For example, a length-type variable could in principle simultaneously determine the width and length of a capacitance pad.

\subsection{Factorize the update step for independent design variables}
The nonlinear minimization step is simplified by noting that the parameters $f_{res}$ and $E_c$ only depend on $l_{res}$ and $w_{qb}$, respectively. Hence, we can reduce the dimension of the minimization problem by running Eq.~\ref{cost} first for the one-dimensional problems $(f_{res}, l_{res})$, $(\kappa_{res}, l_{res-tl})$ and $(f_{qb}, w_{qb})$ to obtain $l_{res}^{k+1}$ and $w_{qb}^{k+1}$ and then minimize the remaining (two-dimensional) problem for $(f_{qb}, \chi, L_{qb}, w_{res-qb})$. In this way we have run the smaller problem of dimensions 1, 1, 1, and 2 instead of running the full 5 dimensional problem, which generally would take longer to solve. As an example, if we define the $l_{res-tl}$ coupling length such that it does not affect the total length of the resonator, we (approximately) decouple the optimization of $f_{res}$ and $\kappa_{res}$.

\subsection{Make effects of design variables predictable}
Depending on how a design variable affects a design, it might have a simpler or more complicated (even non-monotonic) relationship to its parameter. It is generally advisable to keep the relations between design variables and parameters as simple as possible, and to ensure that the chosen design variables are the layout properties that affect each parameter the most when changed. For example, the coupling capacitance between a qubit and a resonator scales linearly with the width of its coupling pad only when the pad is much wider than the resonator’s center conductor.

\subsection{Investigate unknown physical relationships}
In the case where it is difficult to construct a predictable design, the QDesignOptimizer can also be used to investigate the physical relationship between design variables and parameters, which is useful when the relationship is unknown or needs to be verified. To do this, the optimization is run without targets and the design variable of interest is given explicit values in each iteration, see the documentation \cite{qdesignoptimizer_git} for details. The obtained trajectory can then be fitted to a simple model, such as a power law, to obtain an approximate physical relation.

\subsection{Faster convergence}
To reduce the total simulation time, the convergence criteria (for example the number of passes in the HFSS simulation) can be successively increased for each iteration, such that a more detailed simulation only is run when the design variables and parameters are close to target. 

\subsection{Tracking of design variable updates}
\begin{figure}
    \centering
    \includegraphics{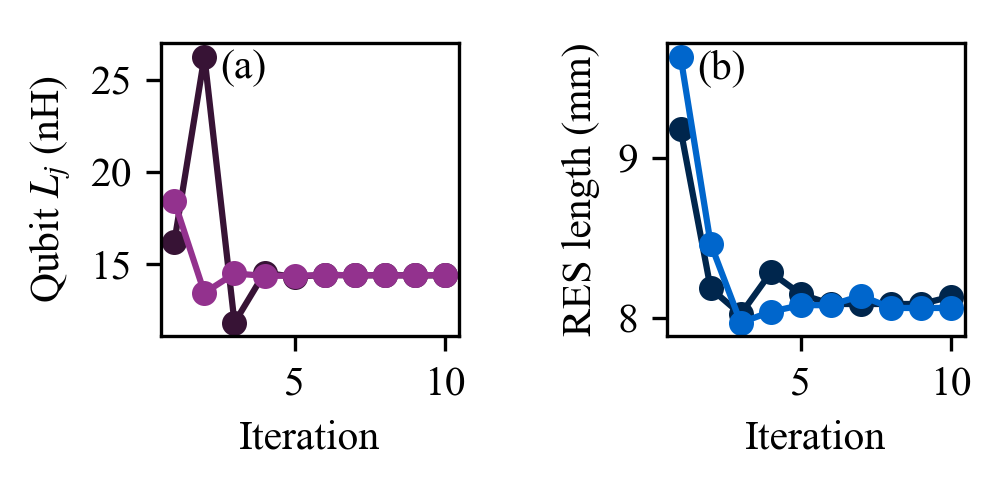}
    \caption{Tracking of two out of five design variables versus iterations for two different sets of initial conditions. (a) Qubit Josephson inductance. (b) Resonator length. The colors of the design variables match the colors of the target parameters in Fig.~\ref{fig:fig2_OPT_single_qubit_resonator}.}
    \label{fig:FigS3_OPT_single_qubit_resonator_tracking_des_var}
\end{figure}

To aid the debugging and help understanding parameter updates within the large design space, the evolution of the design variables through the optimization can also be tracked, as shown in Fig.~\ref{fig:FigS3_OPT_single_qubit_resonator_tracking_des_var} for the case of the qubit Josephson inductance and the resonator length for two different sets of initial conditions. The final design variable values correspond to the parameters close to the desired parameter targets. 

\bibliography{referenes_QDO}

\end{document}